\documentclass[]{aastex62}
\usepackage{multirow}

\shortauthors{Luchsinger, Chanover, and Strycker}

\begin{document}

\title{Water Within a Permanently Shadowed Lunar Crater: Further LCROSS Modeling and Analysis}

\correspondingauthor{Kristen Luchsinger}
\email{kluchsin@nmsu.edu
}

\author{Kristen M. Luchsinger}
\affiliation{Astronomy Department, New Mexico State University, Las Cruces, NM, USA} 
\nocollaboration

\author{Nancy J. Chanover}
\affiliation{Astronomy Department, New Mexico State University, Las Cruces, NM, USA}
\nocollaboration

\author{Paul D. Strycker}
\affiliation{Concordia University Wisconsin, Mequon, WI, USA}
\nocollaboration

\begin{abstract}

The 2009 Lunar CRater Observation and Sensing Satellite (LCROSS) impact mission detected water ice absorption using spectroscopic observations of the impact-generated debris plume taken by the Shepherding Spacecraft, confirming an existing hypothesis regarding the existence of water ice in permanently shadowed regions within Cabeus crater. Ground-based observations in support of the mission were able to further constrain the mass of the debris plume and the concentration of the water ice ejected during the impact. In this work, we explore additional constraints on the initial conditions of the pre-impact lunar sediment required in order to produce a plume model that is consistent with the ground-based observations. We match the observed debris plume lightcurve using a layer of dirty ice with an ice concentration that increases with depth, a layer of pure regolith, and a layer of material at about 6 meters below the lunar surface that would otherwise have been visible in the plume but has a high enough tensile strength to resist excavation. Among a few possible materials, a mixture of regolith and ice with a sufficiently high ice concentration could plausibly produce such a behavior. The vertical albedo profiles used in the best fit model allows us to calculate a pre-impact mass of water ice within Cabeus crater of $5 \pm 3.0 \times 10^{11}$ kg and a mass concentration of water in the lunar sediment of $8.2 \pm 0.001$ \%wt, assuming a water ice albedo of 0.8 and a lunar regolith density of 1.5 g cm$^{-3}$, or a mass concentration of water of $4.3 \pm 0.01$ \%wt, assuming a lunar regolith density of 3.0. These models fit to ground-based observations result in derived masses of regolith and water ice within the debris plume that are consistent with \emph{in situ} measurements, with a model debris plume ice mass of 108 kg. \\
\end{abstract}


\section{Introduction}

The objective of the 2009 Lunar CRater Observation and Sensing Satellite (LCROSS) mission was to detect water ice within a permanently shadowed lunar crater \citep{Colaprete2010}. Because permanently shadowed regions (PSRs) are shielded from solar radiation, water ice could theoretically survive in these regions over geologic time scales \citep{Watson1961,Arnold1979,Paige2010}. The LCROSS involved impacting the spent upper stage of the Centaur rocket originally used to launch the Lunar Reconnaissance Orbiter (LRO) into Cabeus crater, a crater near the southern lunar pole containing PSRs, on October 9, 2009. A Shepherding Spacecraft followed the rocket, taking spectra of the ejected debris, and impacted the lunar surface four minutes after the rocket. The spectra of the debris plume contained signatures of water vapor and of water ice grains, and these were used to determine a value of $5.6 \pm 2.9\%$wt for the concentration by mass of the water within the debris plume material \citep{Heldmann2015, Colaprete2010, Colaprete2012}.\\

Prior to LCROSS, the presence of water ice had been inferred by detections of hydrogen enhancements within PSRs using neutron spectrometers on board the Lunar Prospector and the Lunar Reconnaissance Orbiter (LRO)  \citep{Feldman1998, Feldman2001, Mitrofanov2010}. An absorption band attributed to either hydroxyl (OH) or water ($\rm H_{2}$O) was also observed in spectra obtained by the Visual and Infrared Mapping Spectrometer (VIMS) on Cassini \citep{Clark2009}. However, measurements of the circular polarization ratio did not detect evidence of thick deposits of ice \citep{Stacy1997, Campbell2006}. Thus the detection of water ice and vapor within the LCROSS debris plume represented an important step forward in understanding the formation and evolution of the Moon and, potentially, even of the Earth.\\

In addition to the LCROSS impact, additional techniques were developed to detect and characterize water ice on the Moon. A diurnally varying OH/$\rm H_{2}$O signature on the lunar surface was detected by the Moon Mineralogy Mapper ($\rm M^{3}$) instrument on the Chandrayaan-1 mission and by the Deep Impact spacecraft \citep{Pieters2009, McCord2011, Sunshine2009}. Albedo enhancements detected within PSRs using the Lunar Orbiter Laser Altimeter (LOLA) instrument on board LRO provided additional evidence for water ice within PSRs \citep{Hayne2015, Fisher2017}. Recently, the spectral signature of water ice exposed on the surface of shadowed regions was detected using indirect lighting \citep{Li2018}. Despite these numerous advances in the characterization of the lunar polar environment, however, the LCROSS detection remains unique among these water ice detections due to the fact that the impact excavated and lofted PSR material from below the surface. \\

During the impact of the Centaur rocket, several teams of observers used ground-based telescopes, one of which was the Astrophysical Research Consortium (ARC) 3.5~m telescope at Apache Point Observatory (APO) in Sunspot, NM, to provide simultaneous observations of the ejecta plume \citep{Heldmann2012}. All teams were initially unsuccessful, with no ground-based imaging detections of the plume itself \citep{Chanover2011,Heldmann2012}, although \cite{Killen2010} detected emission from sodium lofted into the Moon's exosphere by the debris plume. However, further analysis of the APO optical images using principal component analysis (PCA) techniques revealed the presence of a plume lightcurve within the APO observations \citep{Strycker2013}. This detection was used to constrain the total mass of the debris plume, and using the total mass of water detected by \cite{Colaprete2010}, \cite{Strycker2013} calculated a concentration of water ice within the debris plume of $6.3 \pm 1.6 \%$wt.\\

Previous modeling efforts motivated by the ground-based detection constrained the structure of the debris plume \citep{Strycker2013}. In order to simulate the lunar conditions that produced the observed debris plume, \cite{Strycker2013} relied on experiments conducted at the NASA Ames Vertical Gun Range that provided constraints on the kinetic properties of the LCROSS debris plume \citep{Hermalyn2012, Schultz2010}. However, the composition of the pre-impact lunar sediment can also affect the observed brightness of the plume as a function of time. The composition of the lunar sediment influences the albedo, or reflectivity, of the particles. Therefore, any variation in composition as a function of depth will translate into a variation of reflectivity of the particles as a function of time, which will therefore affect the shape of the observed lightcurve resulting from sunlight reflecting off of the plume particles. In this study, we explore the previously unexamined parameter space of composition variations within the pre-impact lunar sediment. \\

In \S 2, we present the observations and the data reduction techniques used for this analysis. We discuss our modeling process and present a justification of our choices of model parameters in \S 3. We present a family of models fit to the observations in \S 4. In \S 5, we discuss the results of the models and calculate the corresponding mass concentrations of water ice within the pre-impact lunar sediment within Cabeus crater. We also derive the total mass, water ice mass, and water ice concentration of the debris plume itself. Finally, in \S 6, we present our conclusions and discuss the implications for the delivery source of the water ice within Cabeus crater. \\

\section{Observations}

The ground-based observations analyzed for this study were taken during the LCROSS impact using the Agile camera on the ARC 3.5 m telescope at APO \citep{Chanover2011}. Agile is a high speed time series CCD-based optical photometer with a minimum exposure time of 0.5 seconds \citep{Mukadam2011}. The images were taken in 2$\times$2 binned mode with a pixel scale of 0.26'', which translates to 0.46 km per pixel at the Moon. \cite{Chanover2011} used a ``lunagraph'', a dark slide partially obscuring the illuminated disk of the Moon, to reduce scattered light from the lunar disk. During observations, the atmospheric seeing ranged from 0.8$''$ to 1.4$''$. The observations were smoothed using a 2.5 second boxcar average. \\

A brightening followed by a slow dimming due to scattered light from the debris plume was expected, but initially not detected in the Agile observations using standard image processing techniques \citep{Chanover2011}. Subsequent application of principal component analysis (PCA) techniques revealed the signature of the LCROSS debris plume lightcurve \citep{Strycker2013}. The PCA process identifies principal components (PCs) of the image, given as eigenvectors, that vary over time according to a time series of eigenvalues. By removing the first several PCs, including those that correspond to the lunar landscape, temporal variations in the atmospheric seeing, and coalignment errors, the signal-to-noise ratio (SNR) of lightcurves that would otherwise have been undetectable can increase. This process of PCA filtering allowed \cite{Strycker2013} to achieve a sufficiently high SNR to detect the debris plume.\\

Recent refinement of the PCA filtering pipeline yielded a detection of the debris plume with improved SNR when compared to the images used in \cite{Strycker2013}. The observations were re-reduced, including a field of view that was approximately eight times larger, and more principal components were removed from the images \citep{Schotte2017}. Averaged across 140 pixels covering the central debris plume over post-impact times from 17 to 27 seconds, the re-reduced data had a single-pixel SNR of 6.2$\sigma$, an improvement over the data used in 2013, which had a single-pixel SNR of 3.9$\sigma$. This now enables analysis of the images with a higher spatial resolution compared to the 4$\times$4 binning used in the previous analysis.\\ 

\begin{figure}[h]
    \centering
    \includegraphics[width=0.9\textwidth]{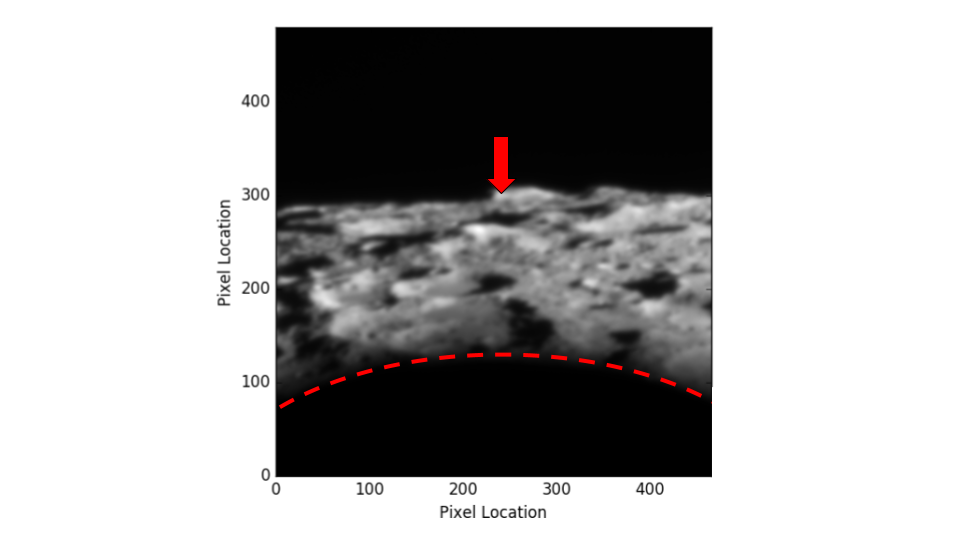}
    \caption{An Agile image taken 25.5 seconds after impact with no PCA filtering, with a red arrow indicating the location of Cabeus crater. The dark curve in the lower half of the image, indicated by the dashed red line, is the ``lunagraph'' used by \cite{Chanover2011} to reduce scattered light from the disk of the Moon. The lunar landscape and the ground-based atmospheric seeing dominate the scene, and the plume is not discernible.}
    \label{f:background}
\end{figure}

\begin{figure}[h]
    \centering
    \includegraphics[width=0.9\textwidth]{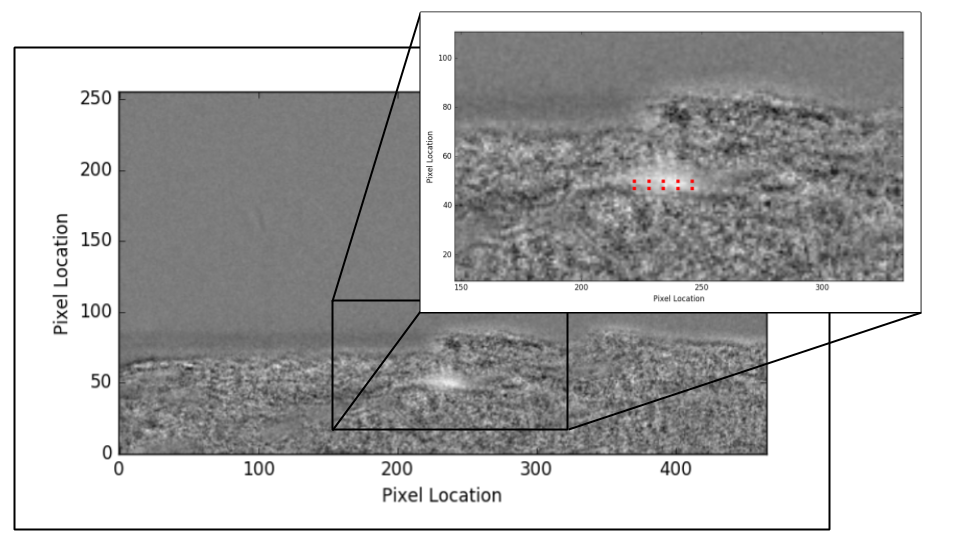}
    \caption{The same Agile image as in Figure \ref{f:background} but with the first 46 PCs removed, using the process described in \cite{Strycker2013}. The PCA-filtered images are cropped, with the bottom half of the image removed in order to avoid the defocused region at the edge of the ``lunagraph''. Overlaid is a zoomed-in view of the debris plume, with the ten red squares indicating the ten physical locations analyzed for this study.}
    \label{f:data}
\end{figure}

\begin{figure}[h]
    \centering
    \includegraphics[width=1.0\textwidth]{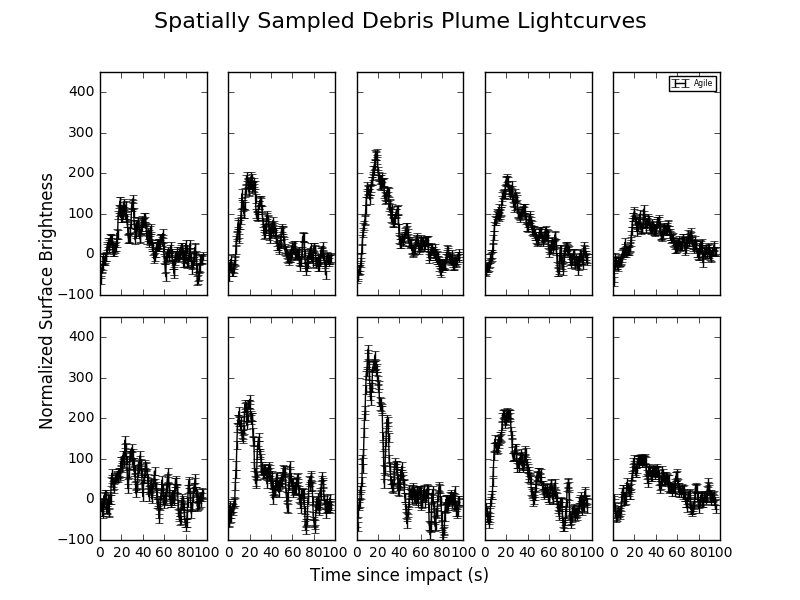}
    \caption{The lightcurves extracted from the ten locations marked with red boxes in Figure \ref{f:data}. }
    \label{f:lightcurves}
\end{figure}

Examples of the Agile images used for this study are provided in Figures \ref{f:background} and \ref{f:data}. Figure \ref{f:background} shows an Agile image acquired 25.5 seconds after the impact, when the debris plume was brightest, before PCA filtering. The plume is not evident in this image. Figure \ref{f:data} shows the same image as in Figure \ref{f:background} but with the first 46 principal components removed, including the static background and the temporal variations in the atmospheric seeing. Figure \ref{f:data} also includes a zoomed in view of the plume, with the locations analyzed in this study indicated by red squares. We extracted the intensity from each of the ten marked locations in the PCA filtered images. The images were taken with a frequency of one every 2.5 seconds, and we extracted intensities from each image between 0 and 96.1 seconds after the impact to produce lightcurves for each of the marked locations, as shown in Figure \ref{f:lightcurves}.\\


\section{Modeling Description}

In order to interpret the observations of the LCROSS ejecta plume, we developed a 141,000 particle three dimensional dynamical model to simulate the debris plume resulting from the impact. All particles were initialized with a velocity and ejection angle, and tagged with a radius `R' and albedo `A'. We treated the outer edge of the Moon as a flat surface, and we assumed that particle-particle and particle-gas interactions were negligible, as in \cite{Bernardoni2019}. We chose initial conditions in order to match the velocity and angle distribution observed experimentally by \citet{Hermalyn2012}; the model was not developed from physical first principals, but rather designed to mimic observations. We therefore did not include particle-particle or particle-gas interactions, as these interactions were present in the experiments, and therefore have already affected the velocities and angles that produce a plume similar to that produced by the experiments. Each particle therefore moved ballistically under the influence of the gravity of the Moon, with its trajectory determined solely by its initial velocity and angle of motion, using the equations of motion under constant acceleration. The ground-based observations are limited to the first six kilometers above the surface and the first ninety seconds after the impact, and therefore we are only sensitive to the early stages of plume evolution. We produced FITS files every 0.1 seconds of the summed reflecting surface area, defined as $\rm A\pi R^{2}$, of all particles within a given pixel, with the FITS image pixels scaled to the pixel resolution of the Agile camera.\\


Using this plume simulation code, we explored variations in the lunar sediment composition in order to compare the simulated plume to the observed plume in an attempt to understand the stratification of the pre-impact lunar sediment. We compared the brightness of pixels from the observations to the pixels at the same physical locations in the time series of simulated images. As illustrated in Figure \ref{f:data}, we compared two rows of five pixels at heights of 3.4 km and 5.3 km above the crater floor. The five pixels in each row sampled the plume at horizontal intervals of 3.8 km. \\

We convolved each simulated image from our model with a 2D Gaussian with a standard deviation corresponding to the atmospheric seeing at time steps equal to the sampling frequency of the Agile images, i.e., one image every 0.5 seconds. We then applied a 2.5 boxcar average to smooth the models to match the observations. We computed the reflectance, defined for the particles in our synthetic images as the surface area of the particle times the albedo of the particle, of all of the particles within each of the model pixels. We compared this to the intensity of light from each of the corresponding pixels in the observations at each time step. We assumed that the model plume is optically thin, and that each particle of a given albedo contributed an equal amount of reflected sunlight. This process produced a time series of the plume brightness, in the case of the observations, or summed particle reflectance, in the case of the synthetic images, within each of the ten analyzed locations.\\

\subsection{Motivation for Further Modeling}

In order to match the observed lightcurve of the LCROSS debris plume, the \cite{Strycker2013} model required a truncated initial velocity distribution, which included no particles ejected at or below 150 m $\rm s^{-1}$. This truncated initial velocity distribution is an altered version of the log-linear initial velocity function for a small scale hollow impactor measured by \cite{Hermalyn2012}. \cite{Strycker2013} did not explore a physical reason for the truncated velocity distribution, although the authors noted that initial velocity is loosely correlated with the initial depth from which the particles were ejected, and therefore a truncated initial velocity distribution may be indicative of an altered albedo with depth.\\

\begin{figure}[h]
    \centering
    \includegraphics[width=1.0\textwidth]{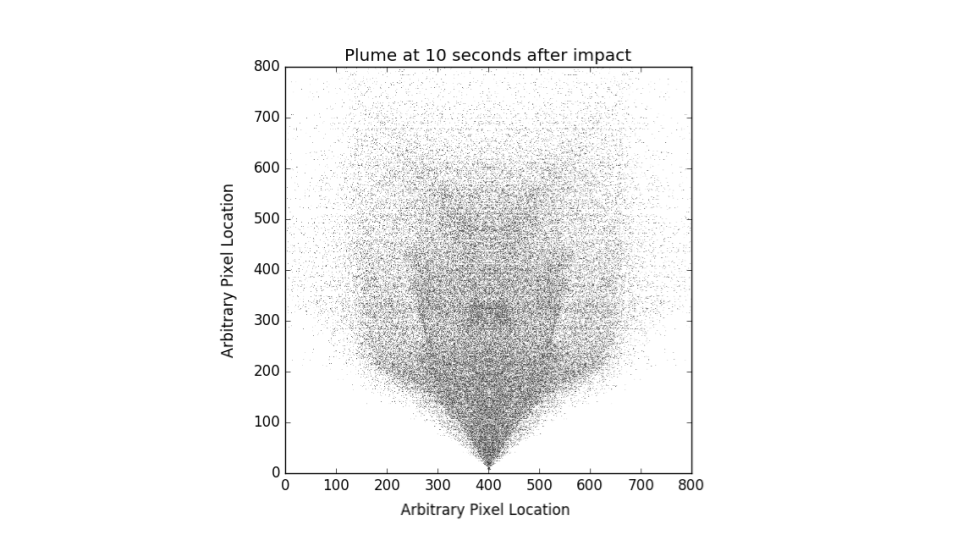}
    \caption{A three-component debris plume model generated by the N-body simulation code used in this work. The plume in this figure was generated with an overdensity of particles in the high and very high angle components in order to demonstrate the angles involved in each component.}
    \label{f:mock}
\end{figure}

\cite{Strycker2013} demonstrated that the modeled plume required three components: a low angle component with angles $\geq 35^{\circ}$, a high angle component with angles between $55^{\circ}$--$75^{\circ}$, and a very high angle component with angles between $75^{\circ}$--$90^{\circ}$ in order to replicate the observed light curve. This three-component plume was consistent with predictions based on the experiments done by \cite{Hermalyn2012}. Figure \ref{f:mock} depicts the three components, with an overdensity of particles in the high angle and very high angle components for improved visibility. The low angle component contains 77.4\% of the total number of particles, the high angle component contains 18.8\%, and the very high angle component contains 3.8\% of the particles, which are the same component percentages as in \cite{Strycker2013}. The triple component plume structure matches the observations better than the single component plume expected from a solid impactor, supporting the ability of the model to accurately probe the shape of the plume.\\
 
The \cite{Strycker2013} truncated initial velocity model fit the images reduced by the original pipeline, but with the new data reduction pipeline we can now analyze the images at finer spatial resolution and at a greater horizontal distance from the center of the plume. At these greater horizontal distances, which were not included in the previous analysis, the truncated initial velocity model no longer matches the observations. In this work, we fit the LCROSS debris plume observations using a stratified lunar sediment model rather than a truncated initial velocity distribution model.\\

\begin{figure}[h]
    \centering
    \includegraphics[width=1.0\textwidth]{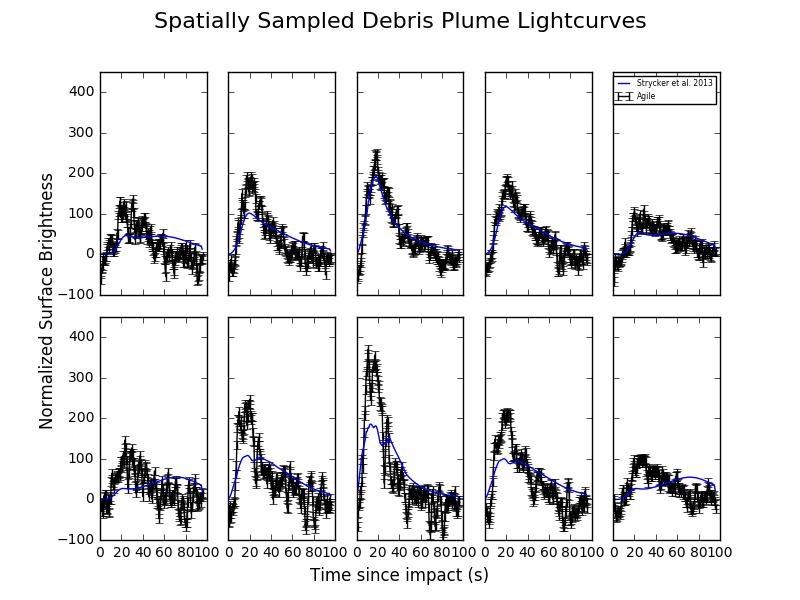}
    \caption{A reproduction of the \cite{Strycker2013} truncated velocity distribution model fit to the re-reduced Agile observations used in this work (the black line), spatially sampling the plume in the 10 discrete regions indicated in Figure \ref{f:data}. The truncated velocity distribution model does not match the intensity of both the central and edge locations of the observed plume, and for the lower altitude pixels, the lightcurve decays more slowly for the synthetic plume than for the observations, resulting in an overintensity between 30 and 80 seconds.}
    \label{f:2013}
\end{figure}
 
As a first step, we reproduced the model used in \cite{Strycker2013}, which used a truncated initial velocity distribution and a constant albedo with depth, and compared this model to the observations reduced by the improved pipeline.  We also explored the parameter space resulting from changing the exponents that alter the upper and lower boundaries of the initial velocity distribution. We found that the best fit to the observations occurred when we used a lower exponent of 0.0, that is, no lower velocity boundary, and an upper exponent of -3.15. The lightcurves produced by this model, shown in blue in Figure \ref{f:2013}, do not match the surface brightness for both the edge and central locations of the observation images and include an overintensity during the decay of the lightcurve between 20s and 50s for the lower row of pixels. We therefore next developed models that explored a different parameter space: the vertical stratification of the pre-impact lunar sediment.\\
 
\subsection{Stratified Lunar Sediment Models}
\label{layers}

In order to explore the parameter space of vertically stratified pre-impact lunar sediment, we held the kinetic behavior of the plume fixed to the distributions measured by \cite{Hermalyn2012}. We developed models of the lunar sediment that include some or all of the following layers: a layer of bedrock, a layer of lunar regolith, a surface layer of a mixture of regolith and water ice, i.e., ``dirty ice'', and a mixing region in which the concentration of water ice decreases with depth.\\

Each of the aforementioned layers is physically motivated by previous observations. A layer of bedrock covered by a layer of lunar regolith is supported by prior optical imaging with the Lunar Reconnaissance Orbiter Camera (LROC) of non-shadowed lunar craters \citep{Domingue2018, Fa2010}. LCROSS impacted with a velocity of 2.5 km $\rm s^{-1}$ and a mass of 2,305 kg \citep{Schultz2010}. The resulting kinetic energy delivered to the crater should have excavated material down to a depth of 5 meters; if material from a depth shallower than 5 meters was not ejected, it is likely a material with a high tensile strength, such as bedrock. We note that while the term ``bedrock'' implies a material with a high tensile strength, other materials may also fit this description, possibly including mare deposits or a cold regolith with temperature less than 100 K mixed with an ice concentration greater than 8 \%wt, which has a hardness approaching that of concrete. We refer to this layer as a ``bedrock layer'', but bedrock should be understood here to describe only the high tensile strength, and should not necessarily be interpreted as a true lunar bedrock. In this model, we treat any material from a ``bedrock'' layer as effectively having an albedo of zero.\\

A layer of surface ``dirty ice'' is supported by measurements of higher albedo within PSRs compared to the albedos of non-shadowed polar regions. The higher albedo is associated with water ice deposits \citep{Lucey2014, Hayne2015, Fisher2017}. This water would have been deposited at the surface of the crater, either delivered by the impactor that created the crater, or implanted on the lunar surface by the solar wind and subsequent migration by the water molecules across the surface until captured by the cold trap of the PSR. Both of these methods would deposit the water onto the surface of the crater, creating a layer of ``dirty ice'', that is, lunar regolith with a higher concentration of water ice. Radar observations of Cabeus crater pre- and post-impact did not provide definitive evidence of a thick surface layer of pure water ice. Some studies have found that the polarization measurements were not inconsistent with discrete water ice particles mixed with lunar regolith \citep{Neish2011}, while others detected a possible water ice signature using the Mini-RF radar instrument on board LRO \citep{Patterson2017}. In our model, we assume that any water ice is in the form of discrete water ice particles mixed with lunar regolith, and we assume that there is no thick surface layer of ice.\\

We also include layers containing vertical gradients in water ice concentration due to regular impacts of micrometeorites. These small impacts disrupt the lunar regolith, allowing for mixing of the material. We therefore also include mixing regions between the surface ice and pure regolith layers in some of our models, in which the concentration of water ice either decreases or increases with depth \citep{Needham2017, Rubanenko2019}. A layer with increasing concentration with depth could be consistent with water sources that have not recently deposited water near the surface, such as water deposited during a period of high volcanic activity or water endogenic to the lunar materials; such a layer could also be consistent with water deposits that have been disturbed by meteorite impact gardening or covered by ejecta blankets from nearby impacts \citep{Hurley2012}. Such a process could perturb the ice concentration at depths of a meter or more \citep{Costello2019}. \\

\subsection{Water Ice Grain Survivability}

In this work, we assume that the bulk of the water present in the lunar sediment is excavated as water ice grains, as opposed to being vaporized during the impact. This is consistent with the spectral results reported in \cite{Colaprete2010}. Figure 3 from \cite{Colaprete2010} shows the spectral signatures of water ice, water vapor, and OH vapor detected at three time epochs just after the LCROSS impact. Between 0 and 23 seconds, all three signatures are detected. Between 23 and 30 seconds, the water vapor and OH vapor signatures have almost entirely disappeared, while the water ice grain signature continues to absorb at roughly the same absorption band depth. Finally, between 123 and 180 seconds, the water ice grain signature disappears and the water vapor signature reappears with a deeper absorption signature than it had in the first 23 seconds. The OH absorption signature does not reappear; however, OH emission is detected with an initial peak at 30 seconds after impact and a continued rise after 120 seconds. \\

These spectroscopic results suggest the formation of an initial water vapor plume immediately after impact. This agrees with thermal modeling by \cite{Stopar2018}, which found that only the first $\sim$3 cm of water ice would be vaporized during the impact. The water vapor was followed by water ice grains excavated in solid form and rising along with the regolith in a longer lasting plume. The water ice grains, once exposed to sunlight, sublimated, leading to the replacement of the water ice grain signature with a new water vapor signature. This process began 30 seconds after impact and lasted for the duration of the observations by the Shepherding Spacecraft \citep{Colaprete2010}. Additionally, \cite{Poondla2020} found that the lifetime of 1 $\mu$m pure ice particles is on the order of $10^{3} - 10^{4}$ seconds. Our observations were acquired between 0.0 and 96.1 seconds after impact, and we see reflected light from the debris plume between 0.0 and 35 seconds. We therefore do not expect a significant amount of water ice grains to have been vaporized during our observations of the debris plume.\\

\subsection{Model Parameters}

We created synthetic plumes using a numerical N-body simulation with a total of 141,000 particles. Our simulation assumes an azimuthally symmetric plume, which allowed us to model one $36^{\circ}$ segment of the plume and populate the rest of the plume with duplicates of the segment. We used a velocity distribution that is linear in log space, following the velocity distribution empirically measured in the \cite{Hermalyn2012} experiments. Our distribution includes velocities well below the truncation lower limit of 150 m $\rm s^{-1}$ used in the \cite{Strycker2013} model, with velocities ranging between 10 m $\rm s^{-1}$ and 700 m $\rm s^{-1}$. We also used the best fit distribution of ejection angles as determined by \cite{Strycker2013}. This best fit distribution includes a lower bound ejection angle for the lower angle component of $35 \pm 5^{\circ}$, compared to the $45^{\circ}$ measured by the \cite{Hermalyn2012} experiments, suggestive of a lower density impacted material.\\ 

In each model, we adopted a nominal lunar regolith albedo value of 0.17, and assumed that this value is constant with depth below the lunar surface \citep{Shkuratov1999, Gold1974}. We did not use a lunar regolith albedo that increases with depth, as in \cite{Gold1975}, since those authors concluded that the darker surface albedo was caused by chemical interactions with the solar wind. Since the material in question is in a PSR, it is more shielded from the solar wind, and should not be chemically darkened in the same way as the Apollo core samples. Furthermore, if the surface material was chemically darkened, the darkening would affect only the first 50 cm of material \citep{Gold1974}. As we will discuss in \S \ref{Modeling}, our observations are insensitive to albedo changes within the first meter of material, therefore we ignore the effect of surface darkening of the lunar regolith.\\

We represented  a surface layer of dirty ice with an albedo of 0.33, following the results of \cite{Lucey2014} derived from LOLA observations at 1064 nm. Our observations were made in visible light, as compared to the LOLA NIR observations; however, mixed ice and regolith will still have a higher albedo than a pure regolith mixture in visible light. An icy region albedo of 0.33 is also consistent with \cite{Hayne2015} and \cite{Fisher2017}, which both found lunar icy region albedos of between 0.3 and 0.35. We do note, however, that our choice of an albedo of 0.33 is lower than the 0.35 cutoff used by \cite{Li2018} while filtering for signatures of water ice. These LOLA albedo observations suggest that the albedo of permanently shadowed lunar regions is statistically higher than the lunar regolith albedo outside of permanently shadowed regions, even when controlling for geometric and geological conditions \citep{Lucey2014}.\\

The particles excavated by the LCROSS impact were inferred to have radii on the order of microns to submicrons \citep{Heldmann2015}. Throughout this analysis we assumed a constant particle radius of 2.5 $\mu$m, as in \cite{Strycker2013} and \cite{Colaprete2010}. There is a degeneracy between a change in albedo and a change in particle radius such that an increase of a factor of two in the particle radius mimics a change of a factor of four in particle albedo. We held the particle radius fixed in order to allow us to probe the albedo of layers of lunar sediment. We do not consider the degeneracy between particle radius and albedo to be a source of uncertainty in those models with a constant or increasing water ice concentration with depth. A decreasing vertically stratified particle radius profile, which would mimic the expected decrease in albedo between surface ice and lunar regolith, is not a physically realistic scenario. We would instead expect a constant or increasing vertically stratified regolith particle radius profile, with smaller particles at the surface due to the influence of micrometeorite impacts and larger particles deeper in the vertical profile. Such an increasing vertical profile would mimic an increase in albedo at a given depth within the lunar sediment. The model containing an increasing albedo profile, therefore, is degenerate with an increasing particle radius profile. All profiles are presented as albedo profiles as a function of depth; they should be understood to possibly be composite profiles with both albedo and particle radius varying with depth, but we assume for this analysis a constant radius profile with depth.\\

We generated models with varied lunar sediment conditions by assigning one of the two albedos, 0.17 or 0.33, to each particle depending on the maximum height reached by the particle during excavation. The maximum height reached by the particle was calculated using Equation 1, where $\rm H_{m}$ is the maximum height reached by the particle, $\rm v_{initial}$ is the initial velocity of the particle, $\rm \theta$ is the ejection angle of the particle, and g is the specific gravity of the Moon:
\begin{eqnarray}
\rm H_{m} = \frac {v_{initial}*\sin^{2}(\theta)} {2g}.
\end{eqnarray}
We used this value of maximum height reached by the particles as a proxy for the initial depth of the particle prior to impact. For the translation from maximum height to maximum depth, we used the experimental relation illustrated in Figure 9 of \cite{Hermalyn2012}. Due to the scatter in this relation between maximum height and maximum initial depth, there will be some mixing between the layers. The uncertainties due to the scatter can be interpreted as being a transitional region between the two boundary layers that is on the order of 0.4--0.8 meters thick. Therefore, although our layers are defined by a step function in albedo with respect to maximum height reached by the particles, the actual transition between the regions is not a step function, but rather is a layer with a mixture of particles from the two bordering regions.\\

\section{Modeling Results}
\label{Modeling}

\begin{figure}[h]
    \centering
    \includegraphics[width=1.0\textwidth]{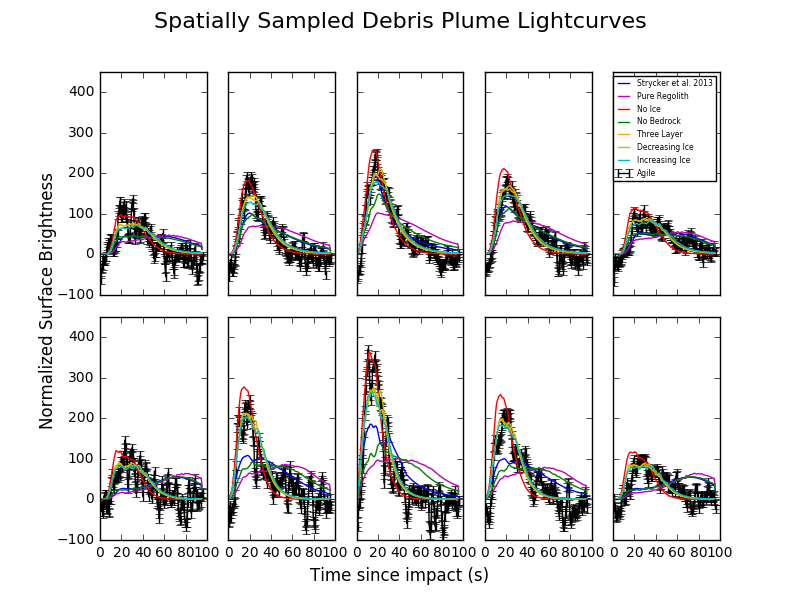}
    \caption{All models fit to the Agile observations (the black line with error bars) spatially sampling the plume in the 10 discrete regions indicated in Figure \ref{f:data}. The blue line is the \cite{Strycker2013} model, the magenta line is the pure regolith model, the red line is the no ice model, the green line is the no bedrock model, the orange line is the three layer model, the yellow line is the Decreasing Ice model, and the cyan line is the Increasing Ice model.}
    \label{f:alla}
\end{figure}
\begin{figure}[h]
    \centering
    \includegraphics[width=1.0\textwidth]{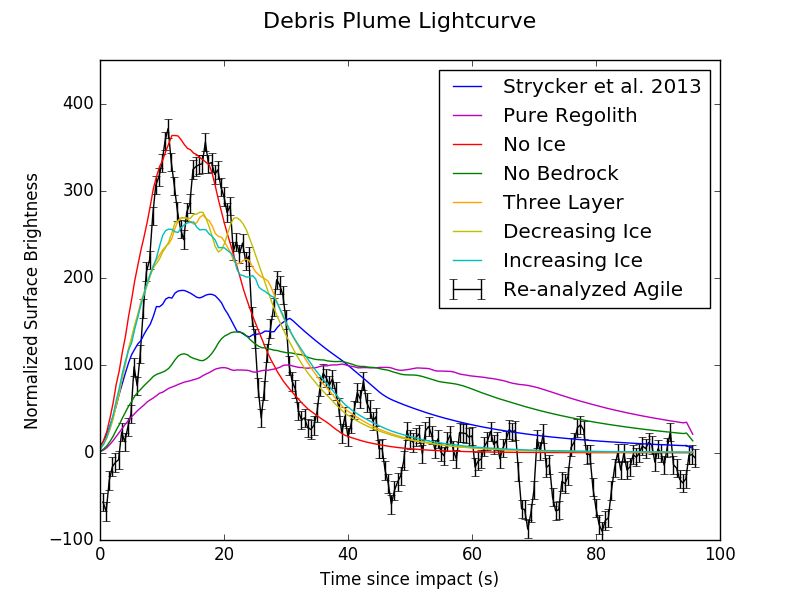}
    \caption{The lower central panel from Figure \ref{f:alla}. All models are fit to the Agile observations (the black line with error bars). The blue line is the \cite{Strycker2013} model. The magenta line is the pure regolith model. The red line is the no ice model. The green line is the no bedrock model. The orange line is the three layer model. The yellow line is the Decreasing Ice model. The cyan line is the Increasing Ice model.}
    \label{f:allone}
\end{figure}

We developed a family of models that explore physically motivated regions of stratified lunar sediment. We first produced lightcurves for a model with a constant albedo, consistent with an impact into pure lunar regolith. This model produced an overintensity during the lightcurve decay in the lower row of pixels between 10 s and 80 s, similar to the overintensity seen in the replicated \cite{Strycker2013} model in Figure \ref{f:2013}. This model is represented by the magenta lines in Figures \ref{f:alla} and \ref{f:allone}. Ultimately, we found that an unstratified lunar sediment profile model does not fit the observations better than the truncated initial velocity distribution model.\\

\begin{table}
\begin{center}
 \begin{tabular}{||c|l|l|l|l|l|l||} 
 \hline
 \multicolumn{7}{||c||}{Model Layers} \\
 \hline
 Depth & Pure Regolith & No Ice & No Bedrock & Three Layer & Decreasing Ice & Increasing Ice\\ 
 \hline
 \multirow{ 4}{*}{$\Bigg\downarrow$} & Regolith & Regolith & Ice & Ice & Ice & Mixing (Increasing) \\ 
 \cline{2-7} 
 & N/A & Bedrock & Regolith & Regolith & Mixing (Decreasing) & Ice \\ 
 \cline{2-7} 
 & N/A & N/A & N/A & Bedrock & Regolith & Bedrock \\ 
 \cline{2-7} 
 & N/A & N/A & N/A & N/A & Bedrock & N/A \\ 
 \hline
 \end{tabular}\\
 \caption{The layers included in each of the models. ``Ice'' refers to a mixture of ice and regolith; ``Mixing (Increasing)'' and ``Mixing (Decreasing)'' refer to layers where the ice concentration either increases or decreases with depth; and ``Bedrock'' refers to a material that is not ejected by the impact, possibly due to high tensile strength.}
 \label{t:layers}
 \end{center}
\end{table}

\begin{figure}[h]
    \centering
    \includegraphics[width=1.0\textwidth]{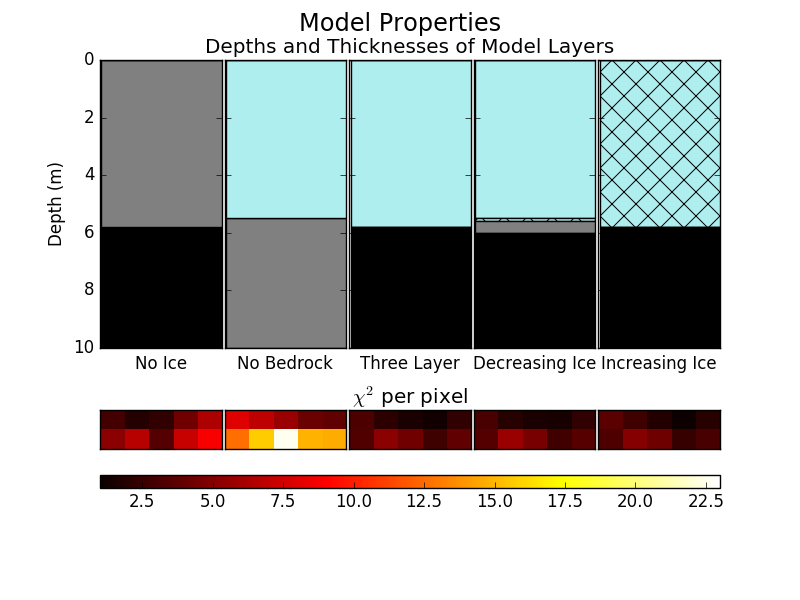}
    \caption{A comparison of the properties of the five stratified lunar sediment models. On the top, we present the best fit thicknesses and locations of the layers in each model. This diagram does not include errors in the depths, and is intended for use only as a visual illustration of the model layers. Light blue represents the surface ice layers; light blue with hatching represents the transitional regions between surface ice and pure regolith; gray represents the pure regolith layers; and black represents the bedrock layers. Below, we present a heat map of the $\chi^{2}$ per pixel for each model. We notice a consistently higher $\chi^{2}$ in the lower left pixels; this is likely due to the asymmetric height of the crater rim, which led to a higher scattered light signal in those pixels, and the increased $\chi^{2}$ effect is constant across all models. We also find that the $\chi^{2}$ is higher for all models in the lower row of pixels, again due to the increase of scattered light near the crater rim. See Figures \ref{f:background} and \ref{f:data} for a visual representation of the slight asymmetry in the crater rim and see Figure \ref{f:alla} for the scattered light asymmetry in the resulting lightcurves.}
    \label{f:notional}
\end{figure}

We next developed four stratified lunar sediment models with different combinations of the physically motivated layers described in \S \ref{layers}: a model with a surface layer of ``dirty ice'' over lunar regolith, i.e., ``No Bedrock''; a model with lunar regolith over bedrock, i.e., ``No Ice''; a model with surface ice, regolith, and bedrock, i.e., ``Three Layer''; and two models with all three layers as well as mixing region layers between the surface ice and regolith layers, i.e., ``Decreasing Ice'' and ``Increasing Ice'', where the regolith layer for Decreasing Ice occurs between the ice and the bedrock, with the ice concentration decreasing with depth, and the regolith layer for Increasing Ice occurs at the surface, with the ice concentration increasing with depth. A list of the layers, ordered in terms of increasing depth, is presented in Table \ref{t:layers}, and an illustration depicting the thicknesses of the layers is presented in Figure \ref{f:notional}. Some layers listed in Table \ref{t:layers} may not be visible in Figure \ref{f:notional} if the model is a better match to the observations when the thickness of that layer goes to zero. This is the case for the Three Layer model; despite containing three possible layers, the model matches the observations best when the pure regolith layer thickness goes to zero.\\

\begin{figure}[h]
    \centering
    \includegraphics[width=1.0\textwidth]{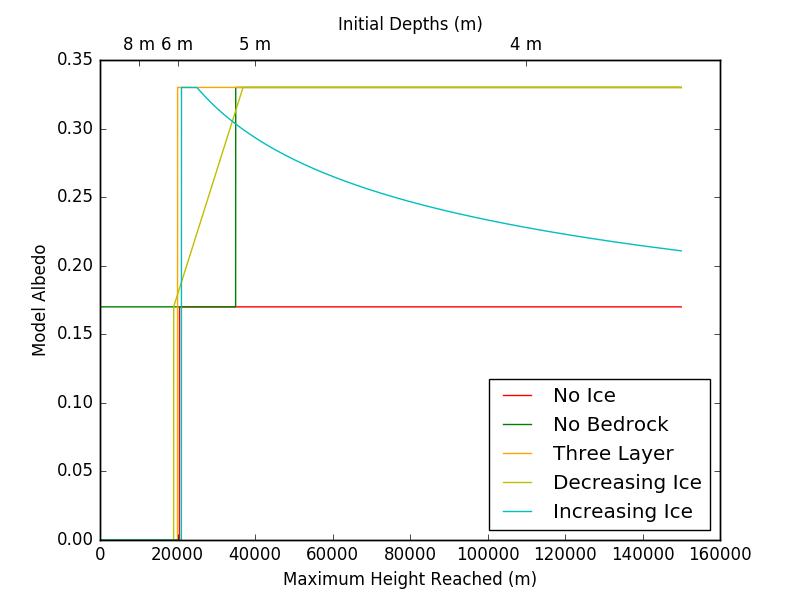}
    \caption{Best fit functions of albedo with respect to depth for four models. The profiles used assigned albedo based on the maximum height reached by the particle, which is then translated into the initial depth below the surface. Three representative depths are included for reference. The red line represents a model with regolith and bedrock, but no ice. The green line represents a model with ice and regolith, but no bedrock. The orange line represents a model with ice, regolith, and bedrock. The yellow line represents a model with ice, regolith, bedrock, and a transition region. The cyan line represents a model with ice, regolith, bedrock, and an increasing ice concentration with depth. The transition to bedrock occurs at 20,000 m for the No Ice, Three Layer, and Increasing Ice models; we have staggered the lines in this plot by 500 meters for visibility.}
    \label{f:albedofuncs}
\end{figure}

The albedo functions used in these models are shown in Figure \ref{f:albedofuncs}. The figure includes three representative depths, illustrating that we are most sensitive to the region of the sediment profile between 4 and 6 meters below the surface. The improved sensitivity between 4 and 6 meters is due to the experimental relation between maximum height and initial depth presented in Figure 9 from \citep{Hermalyn2012}, which is roughly linear in log-log space. The result is that changes in initial depth between 6 meters and 8 meters correlate to small changes in maximum height reached by the particle, while changes in initial depth between 4 m and 6 m correlate to more significant changes in maximum height reached.  We describe the models in terms of maximum height of the particle due to the errors introduced when translating to initial depth; however, we will describe that translation and present the values for initial depth, along with the associated uncertainties, in \S \ref{thickness}. Each of these five models is described in more detail below.\\

\subsection{No Ice Model}
\label{bedrock}

Our first stratified lunar sediment model was motivated by the truncated velocity distribution used in \cite{Strycker2013}. However, whereas \cite{Strycker2013} modified the initial velocity distribution such that no particles would be ejected at a velocity lower than 150 m $\rm s^{-1}$, we set the albedo of all particles that reached a maximum height of less than a cutoff height equal to zero. The zero albedo particles are equivalent to a layer from which no particles were ejected beyond a certain depth, which corresponds to a layer with a sufficiently high tensile strength to resist the force of the impact. This model is consistent with lunar sediment containing no water ice. \\

The no ice model produces lightcurves that do not match the surface brightness of the pixels sampling the edges of the plume in the observations. However, the lightcurve decay in the lower row of pixels produced by this model matches that seen in the observations, unlike the truncated initial velocity and the pure regolith models, which show overintensities during the lightcurve decay. The results of this model are represented as the red lines in Figures \ref{f:alla} and \ref{f:allone}. We calculated a reduced $\chi^{2}$, summed over all pixels, for a variety of maximum heights cutoffs, and found that the minimum summed $\chi^{2}$ occurs when the maximum height cutoff was 20,000 m.

\subsection{No Bedrock Model}
\label{dirtyice}

We also generated a model in which an albedo of 0.33 was assigned to particles that reach a maximum height above a cutoff maximum height, and an albedo of 0.17 was assigned to particles that do not reach that maximum height cutoff. This corresponds to a layer of surface ``dirty ice'' over a layer of pure lunar regolith, and assumes that the lunar regolith layer is sufficiently thick that no bedrock material is reached by the impact. The results of this model are represented as the green lines in Figures \ref{f:alla} and \ref{f:allone}. \\

The no bedrock model produces lightcurves that are very similar to the unstratified, pure regolith model, with a slightly higher peak. For the no bedrock model, we found that the minimum summed $\chi^{2}$ occurs when the maximum height cutoff was 35,000 m. However, the added layer of higher albedo material alone does not add enough reflectance early in the time series to match the data. A bedrock layer appears to be necessary for a stratified lunar sediment profile to match the data. \\

\subsection{Three Layer Model}
\label{Double}

We next developed a three layer model with a surface layer of dirty ice, a layer of pure regolith, and a layer of bedrock. We used step functions for both of the transitions between the layers, as shown in Figure \ref{f:albedofuncs}. We first found the maximum height of the particles at the boundary between the surface layer of dirty ice and the regolith layer that produced the minimum $\chi^{2}$ while holding the maximum height of the particles at the boundary between the bedrock and regolith fixed, and then we held fixed the surface layer-regolith boundary particles while minimizing the $\chi^{2}$ for the bedrock-regolith boundary. These optimized maximum height cutoffs both occurred at 25,000 m, which means that a dry regolith layer does not improve the fit of the model. This was not a constraint placed on the model; rather, the model results suggest that the three layer model matches the observations best when the dry regolith layer has a thickness of zero meters. The fit of the three layer model to the pixels sampling the edges of the plume is an improvement over the bedrock model. The results of this model are represented by the orange lines in Figures \ref{f:alla} and \ref{f:allone}.\\

\subsection{Transition Region Models}
\label{albedo1}

We then generated a model where the transition between the surface layer of dirty ice and the layer of lunar regolith was a linearly decreasing function instead of a step function, which we call the Decreasing Ice model. We consider this a four layer model with the second layer being a 'mixing region' between the surface ice and the lunar regolith. The mixing region represents a lunar regolith that becomes dryer deeper into the lunar sediment. The mixing region lies between maximum heights of 19,000 and 40,000 meters. The albedo decreases linearly with the maximum height reached by the particle, as seen in Figure \ref{f:albedofuncs}. The Decreasing Ice model also includes a bedrock layer, with a step function transition between regolith and bedrock. The minimum summed $\chi^{2}$ occurs when maximum height that would have been reached by the bedrock particles is 19,000 m, the same maximum height as the lower boundary of the mixing region. Similar to the three layer model, Decreasing Ice produces lightcurves that fit both the initial rise and the lightcurve decay of the observations. The results are represented by the yellow lines in Figures \ref{f:alla} and \ref{f:allone}. \\

Finally, we generated a model with a layer with an ice concentration that increases with depth, which we call Increasing Ice, in an effort to explore the results of the removal of near-surface ice by micrometeorite gardening. The ice concentration in this model goes inversely as the maximum height to the one fourth power reached by the particle, followed by a step function transition between the lunar regolith layer and a bedrock layer. At the surface, the albedo of the ice and regolith combination is 0.22 - lower than the maximum of 0.33, but still higher than a dry regolith albedo of 0.17. The dirty ice layer becomes saturated, with an albedo of 0.33, at 25,000 m. The step function from dirty ice to bedrock also occurs at 20,000 m. The results of this model are represented by the cyan lines in Figure \ref{f:alla}.\\

The Increasing Ice model produces lightcurves that are almost identical to those from the Decreasing Ice and the triple layer models, with only minor differences in shape. We therefore conclude that the observations are not sensitive to the rate at which the water ice concentration changes between the surface ice and pure regolith layers.\\

\subsection{Model Comparisons}
\label{stats}

\begin{table}
\begin{center}
 \begin{tabular}{||l|l|l|l|l|l|l||} 
 \hline
 \multicolumn{7}{||c||}{Minimum Summed $\chi^{2}$} \\
 \hline
 \cite{Strycker2013} & Pure Regolith & No Ice & No Bedrock & Three Layer & Decreasing Ice & Increasing Ice\\ 
 \hline
 83.8 $\pm$ 1.02  & 163.3 $\pm$ 1.02  & 49.4 $\pm$ 1.02  & 109.5 $\pm$ 1.02  & 30.6 $\pm$ 1.02  & 29.9 $\pm$ 1.02  & 29.5 $\pm$ 1.02 \\ 
 \hline
 \end{tabular}\\
 \caption{Minimum total $\chi^{2}$ values for all models considered, summed over ten sampled regions.}
 \label{t:min}
 \end{center}
\end{table}

We calculated a reduced $\chi^{2}$ for ten pixels in each model; the total summed $\chi^{2}$ values for each model are presented in Table \ref{t:min}. We additionally present a heat map showing the individual $\chi^{2}$ values for each pixel for each model in Figure \ref{f:notional}. The model that best fit the Agile observations was the Increasing Ice model, with a total $\chi^{2}$ of 29.5 when summed over all ten sampled regions, for an average $\chi^{2}$ per pixel of 2.95.\\

We calculate a Gaussian uncertainty of our $\chi^{2}$ values of $\rm \sqrt{\frac{2}{N}} = 0.102$ per pixel, or 1.02 total for all ten pixels, where N = 192 is the number of data points. The differences in $\chi^{2}$ between the Three Layer model, the Decreasing Ice model, and the Increasing Ice model are therefore less than the uncertainties in the $\chi^{2}$ measurement, and we cannot say that the observations definitively support one of the three models over the others. We therefore conclude that the data are not sensitive to the behavior of ice near the surface. However, we can determine that the models that include layers of ice and bedrock all have a lower $\chi^{2}$ than the models which are missing one of these layers. The stratified lunar sediment models all fit the data better than the unstratified models, and all models with at least a bedrock layer fit the data better than the truncated initial velocity distribution model.\\

\section{Analysis and Results}
\label{thickness}

\begin{table}
\begin{center}
 \begin{tabular}{||l|l|l|l|l||}  
 \hline
   & No Bedrock & Three Layer & Decreasing Ice & Increasing Ice \\
  \hline
 Surface Region & 5.5 $\pm$ 0.7 m & 5.8 $\pm$ 0.5 m & 5.5 $\pm$ 0.7 m & N/A\\ 
 Boundary & & & &\\ 
 \hline
 Mixing Region & N/A & N/A & 5.6 $\pm$ 0.6 m & 5.6 $\pm$ 0.75 m\\ 
 Boundary & & & &\\ 
  \hline
 Pure Regolith & N/A & 5.8 $\pm$ 0.5 m & 6.0 $\pm$ 0.7 m & N/A \\ 
 Boundary & & & &\\ 
 \hline
  Dirty Ice & N/A & N/A & N/A & 5.8 $\pm$ 0.75 m\\ 
 Boundary & & & &\\ 
 \hline
  \end{tabular}\\
 \caption{Depths of Layer Boundaries}
 \label{t:depths}
\end{center}
\end{table}

In our models, we used the maximum height reached by a particle as the determining initial condition when assigning an albedo to a particle. We then translated the maximum height to a maximum initial depth using the distribution presented in Figure 9 from \cite{Hermalyn2012}, as described above.  We present the maximum excavation depths for the boundaries between the layers of each model in Table \ref{t:depths}. The uncertainties in the maximum depths are due to scatter in the data presented in Figure 9 of \cite{Hermalyn2012}.\\

Those models that include a layer with zero albedo, corresponding to material with a higher tensile strength than the material above it, all match the observations best when the maximum depth of the zero albedo layer is 5.8 $\pm$ 0.5 meters. The final crater produced by the LCROSS impact was predicted to be 5.4 meters deep; however, material from up to ten meters below the surface was predicted to be excavated during the impact \cite{Korycansky2009, Hermalyn2012}. In our observations, each pixel corresponds to a spatial distance of 0.46 km, and the lower row is centered on 3.4 km, so in the pixels sampled in this work we can detect particles that reach a height of at least 3.17 km. Translating a maximum height of 3.17 km into a maximum depth from which the particle was ejected using Figure 9 from \cite{Hermalyn2012}, we can detect particles that were ejected from depths shallower than or equal to $9.5 \pm 0.5$ meters below the lunar surface in the lower row of pixels. Any ejected particles with initial depths between 5.8 $\pm$ 0.5 meters, the depth of the zero albedo layer boundaries in our models, and $9.5 \pm 0.5$ meters would be visible in the ground-based observations.\\

A regolith thickness of 5.8 $\pm$ 0.5 meters would indicate that Cabeus is shallower than the 10-15 meters typical of lunar highlands craters \citep{Fa2010}, perhaps suggesting that Cabeus could be similar to Tsiolkovsky \citep{Domingue2018}. However, the layer we refer to as a ``bedrock layer'' merely refers to a layer of material that is not launched above 3.17 km by the impact despite predictions based on energetics. Other materials than bedrock could also fit the description of what we call a ``bedrock layer''. Some possible materials include a buried mare basalt deposit such as that seen in Tsiolkovsky with high tensile strength, large boulders that are ejected but that are not lofted above 3.17 km due to their larger masses, or cold regolith with a temperature lower than 100 K and an ice concentration by mass higher than 8 \%wt  \citep{Greenhagen2016}. \\

The transitions between the layers occurred at the same initial depths for the surface ice layers as well as for the bedrock layers, with the measured values being well within the error bars of the values for the other models. All models with surface ice layers are consistent with no ice deeper than about 5.6 meters. These similarities in transition locations were not constraints placed on the models.\\ 

\subsection{Volume and Mass of Water Ice in Pre-Impact Lunar Sediment}

Assuming the vertical albedo profiles used to match the simulated plume to the observed plume, we derived the concentration and mass of water ice within the pre-impact lunar sediment. We calculated the concentration of water ice by mass using $A_{M} = \frac {0.17+x*A_{W}}{x+1}$, where 0.17 is the albedo of lunar regolith, $A_{W}$ is the albedo of water ice, $A_{M}$ is the albedo found by the model at a given depth, and $x$ is the fractional volume occupied by water ice relative to lunar regolith. We then used this fractional volume to calculate the total volume of water in cubic meters within a column of lunar sediment below one square meter of lunar surface. Solving for $x$:
\begin{eqnarray}
\rm x = \frac{\frac{0.17}{A_{M}} - 1}{1 - \frac{A_{W}}{A_{M}}}.
\label{e:x}
\end{eqnarray}
We then used the value found for $x$ to determine the volume of water, $V_{W}$, as a fraction of the volume of the surface layer of dirty ice, $V$: 
\begin{eqnarray}
\rm V_{W} = V  \frac{x}{x+1}.
\end{eqnarray}
We also calculated the total mass of water ice represented by this volume. We used 0.934 g cm$^{-3}$ for the density of water ice, which is the density of water at 93 K. The radius of Cabeus crater is 30 km; however, only the northwest portion of Cabeus is permanently shadowed \citep{Kozlova2010}. We therefore assumed a circular crater floor with a surface area of $\rm \pi R^{2}$, where R is 15 km, as an approximation for the surface area of the PSR within Cabeus crater. We also assumed that the location of the LCROSS impact was representative of the sediment across the entirety of the floor of Cabeus crater. We then calculated the total mass and volume of water within the pre-impact sediment throughout Cabeus crater. The results of this calculation for the four models containing water ice layers are given in Tables \ref{t:vol} and \ref{t:mass}.\\

Our method of measuring total water mass, water volume, and mass concentration of water (discussed in \S \ref{conc}) is sensitive to the choice of albedo for pure water ice, $\rm A_{W}$. We elected to use an albedo of 0.8 for water ice; however, a range of albedos between 0.6--1.0 could be justified based on albedo values measured for ice in Antarctica \citep{Warren2006}. We therefore present values of water ice mass, volume, and mass concentration calculated for several values in this range of albedos. When presenting our final results, we adopt the values calculated using a water ice albedo of 0.8, based on observations of icy satellites such as Europa, Enceladus, and comets.\\

\begin{table}
\begin{center}
 \begin{tabular}{||l|l|l|l|l||}   
  \hline
  Albedo of Pure Water Ice & No Bedrock & Three Layer & Decreasing Ice & Increasing Ice \\
   \hline
  0.6 & 1.4 $\pm$ 0.9 & 1.6 $\pm$ 0.6 & 1.4 $\pm$ 1.0  & 0.7 $\pm$ 0.5 \\
 \hline
 0.7 & 1.1 $\pm$ 0.6 & 1.2 $\pm$ 0.5 & 1.2 $\pm$ 0.8 & 0.6 $\pm$ 0.3\\
 \hline
  0.8  & 0.8 $\pm$ 0.5 & 0.9 $\pm$ 0.3 & 0.8 $\pm$ 0.5 & 0.5 $\pm$ 0.3 \\ 
 \hline  
 0.9 & 0.8 $\pm$ 0.6 & 0.9 $\pm$ 0.3 & 0.8 $\pm$ 0.5 & 0.4 $\pm$ 0.4\\
 \hline
 \end{tabular}\\
 \caption{Total Volume of Water Ice within Cabeus crater ($\times 10^{9}$ $\rm m^{3}$)}
 \label{t:vol}
\end{center}
\end{table}

\begin{table}
\begin{center}
 \begin{tabular}{||l|l|l|l|l||}   
  \hline
   Albedo of Pure Water Ice & No Bedrock & Three Layer & Decreasing Ice & Increasing Ice \\
     \hline 
  0.6  & 1.3 $\pm$ 0.9 & 1.4 $\pm$ 0.6 & 1.3 $\pm$ 0.9 & 0.7 $\pm$ 0.4 \\ 
  \hline 
  0.7  & 1.1 $\pm$ 0.6 & 1.2 $\pm$ 0.5 & 1.1 $\pm$ 0.7 & 0.6 $\pm$ 0.3 \\ 
    \hline 
  0.8  & 0.9 $\pm$ 0.5 & 1.0 $\pm$ 0.5 & 0.9 $\pm$ 0.6 & 0.5 $\pm$ 0.3 \\ 
    \hline 
  0.9  & 0.8 $\pm$ 0.5 & 0.8 $\pm$ 0.3 & 0.8 $\pm$ 0.5 & 0.4 $\pm$ 0.2 \\ 
  \hline
 \end{tabular}\\
 \caption{Total Mass of Water Ice within Cabeus crater ($\times 10^{12}$ kg)}
 \label{t:mass}
\end{center}
\end{table}

\subsection{Concentration of Water Ice in Pre-Impact Lunar Sediment}
\label{conc}

Previous investigations placed lower limits on the concentration, or fractional mass, of water measured in the spectra of the ejecta plume of 5.6 $\pm$ 2.9 \%wt \citep{Colaprete2010} and 6.3 $\pm$ 1.6 \%wt \citep{Strycker2013}, respectively. We computed the concentration of water ice within the lunar sediment in Cabeus crater:
\begin{eqnarray}
\rm \frac{M_{W}}{M_{tot}} = \frac{M_{W}}{M_{W} + M_{R}}.
\end{eqnarray}
Here, $M_{R}$ represents the total mass of the lunar regolith, including both the regolith present within the surface dirty ice layer and the pure regolith layer. We used a density of lunar regolith of 1.5 g $\rm cm^{-3}$, motivated by measurements of Apollo samples \citep{Carrier1973}. We also present results calculated using a regolith density of 3.0 g $\rm cm^{-3}$ in Table \ref{t:plume} to facilitate comparison between our results and previous work \citep{Colaprete2010, Strycker2013}. The mass concentration of water ice is dependent on the choice of lunar regolith density, but the choice of lunar regolith has no effect on the total volume or mass of lunar ice values presented in Tables \ref{t:vol} and \ref{t:mass}. \\

\begin{table}
\begin{center}
 \begin{tabular}{||l|l|l|l|l||}  
 \hline
  Albedo of Pure Water Ice & No Bedrock & Three Layer & Decreasing Ice & Increasing Ice \\
  \hline
 0.6 & 13.8 $\pm$ 1.1\% & 26.9 $\pm$ 0.0\% & 26.1 $\pm$ 0.6\% & 12.3 $\pm$ 0.01\% \\ 
 \hline
  0.7 & 11.0 $\pm$ 0.6\% & 21.2 $\pm$ 0.0\% & 20.5 $\pm$ 0.6\% & 9.8 $\pm$ 0.01\%\\ 
 \hline
   0.8 & 9.2 $\pm$ 0.7\% & 17.5 $\pm$ 0.0\% & 16.9 $\pm$ 0.6\% & 8.2 $\pm$ 0.01\%\\ 
 \hline
   0.9 & 7.8 $\pm$ 0.5\% & 14.8 $\pm$ 0.0\% & 14.4 $\pm$ 0.6\% & 7.0 $\pm$ 0.01\%\\ 
 \hline
    1.0 & 6.9 $\pm$ 0.4\% & 12.9 $\pm$ 0.0\% & 12.5 $\pm$ 0.6\% & 6.1 $\pm$ 0.01\%\\ 
 \hline
 \end{tabular}\\
 \caption{Concentration of Water by Mass}
 \label{t:concentration}
 \end{center}
\end{table}

The concentrations calculated using this method are given in Table \ref{t:concentration}. Although our modeled depths are maximum possible depths, the calculation of the concentration of water ice within lunar sediment depends on relative depth, not absolute depth, and is therefore not an upper limit for the models that contain lower boundaries in the form of bedrock layers. In the case of the no bedrock model, we assume the regolith extends down to the deepest expected ejected material, 10 meters below the surface. However, if the regolith extends further down, the water ice concentration would decrease. Therefore, for the no bedrock model, the water ice concentration is an upper limit due to the absence of a bounding bedrock layer. The concentrations are calculated using the relative thicknesses of the surface ice and dry regolith layers. If the uncertainties in the depths of the boundary layers are not equal, the concentrations will also have uncertainties. In the case of the three-layer model, there is no dry regolith layer, and therefore the uncertainty in the depth boundary between the surface ice and bedrock layers does not result in an uncertainty in the water ice concentration for that model.\\

With the exception of the No Bedrock and Increasing Ice models, these values are all higher than the 4--6 \%wt derived from the spectroscopic observations of the plume itself \citep{Colaprete2010}. However, this work assumes a lower lunar regolith density; a more accurate comparison between the concentrations is shown in Table \ref{t:plume}. Additionally, the analysis described in this work is calculating the concentration within the pre-impact sediment. We do not expect all of the pre-impact material to have been visible or spectroscopically identifiable within the field of view of the LCROSS Shepherding Spacecraft, therefore the spectroscopic observations may not be measuring the same concentration as was present in the pre-impact sediment. This is also not the first study of Cabeus crater and the LCROSS impact region to suggest that the data are consistent with a high water ice concentration. Using LRO Diviner data, \cite{Hayne2010} modeled the pre-impact sediment and found that a pre-impact concentration of 15-22 \%wt reproduced the detected amounts of water vapor, and that lower concentrations would not have produced the amount of water vapor observed by the shepherding spacecraft.\\

All of the concentrations we find for our models are consistent with the upper limit of 22 \%wt mass concentration that corresponds to a saturation of the lunar regolith. A concentration higher than 22\% might suggest high levels of frost that had not mixed with the lunar regolith. However, because we expect the first couple of centimeters of water ice to have been vaporized, we would not expect our model to be sensitive to such a frost layer, and therefore do not consider a concentration lower than 22 \%wt to be evidence of a lack of a layer of frost. \citep{Stopar2018}.\\

We present values for water ice mass, volume, and concentration for each of the four models containing layers with water ice. However, the model with only surface ice and no bedrock does not match the observations, and the $\chi^{2}$ statistics for the three layer, Decreasing Ice, and Increasing Ice models do not differ at a statistically significant level. We therefore find that the ground-based observations of the LCROSS impact can be equally well fit by models with water ice mass concentrations ranging between 4.95 $\pm$ 0.01 \%wt and 9.6 $\pm$ 0.0 \%wt, calculated using a regolith density of 3 g cm$^{-3}$. The absence of strong evidence in favor of the mixing region layers used in the transition region models may support the model with fewer components, that is, the three layer model. However, the effects of impact gardening seen in the Increasing Ice model are physically motivated, and the mass concentration value for the Increasing Ice model agrees with the mass concentration value calculated by \cite{Colaprete2010}. We therefore consider the Increasing Ice model to be the most likely of the three models. However, we also consider a lunar regolith density of 1.5 g $\rm cm^{-3}$ to be more accurate than a value of 3.0 g $\rm cm^{-3}$; therefore, we present the value of $8.2 \pm 0.01$\%, the value for the Increasing Ice model at 0.8 water ice albedo and 1.5 g $\rm cm^{-3}$ lunar regolith density, to be the most likely water ice concentration value found by fitting stratified lunar sediment models to the ground-based observations of the LCROSS impact.\\

\subsection{Masses of Water Ice and Regolith in the LCROSS Debris Plume}
\label{massplume}

In addition to deriving the mass concentration of water, total mass of water, and total volume of water in pre-impact sediment using the vertical albedo profiles, we also followed the \cite{Strycker2013} method to compute the mass of the debris plume from the scaling used to match the brightness of the model plume signal with that of the Agile data. For direct comparison with their results and with those of \cite{Colaprete2010} we used the same density of lunar regolith and particle radius, 3.0 g $\rm cm^{-3}$ and 2.5 $\mu$m, respectively. Our application of their method needed to allow for more than one particle albedo, $A_{M}$, in the model plume. Therefore, before deriving the mass contribution from each given $A_{M}$ in our model, we calculated its average density:
\begin{eqnarray}
\rm \rho_{A_{M}} = \rho_{W}\left(\frac{x}{x+1} \right) + \rho_{R}\left(1-\frac{x}{x+1}\right),
\end{eqnarray}
where $\rho_{W}=0.934$ g $\rm cm^{-3}$ is the density of pure water ice and $x$ is defined in Eq. \ref{e:x}.\\

\begin{table}
\begin{center}
 \begin{tabular}{||l|l|l|l|l|l||}  
 \hline
  Model & Regolith  & Mass of & Mass of & Total Mass of & Concentration \\
   & Density & Water Ice (kg) & Regolith (kg) & Debris Plume (kg) & of Water by Mass \\
 \hline
  Three Layer & 1.5 g $\rm cm^{-3}$ & 267 & 1263 & 1530 & 17.5\% $\pm$ 0.0\%\\
  & 3.0 g $\rm cm^{-3}$ &  & 2526 & 2793 & 9.6 $\pm$ 0.0\%\\
  \hline
  Decreasing Ice & 1.5 g $\rm cm^{-3}$ & 264 & 1293 & 1555 & 16.9\% $\pm$ 0.6\%\\
   & 3.0 g $\rm cm^{-3}$ &  & 2598 & 2862 & 9.2 $\pm$ 0.6\%\\
 \hline
  Increasing Ice & 1.5 g $\rm cm^{-3}$ & 108 & 1094 & 1202 & 8.2\% $\pm$ 0.01\%\\
  & 3.0 g $\rm cm^{-3}$ &  & 2193 & 2301 & 4.3 $\pm$ 0.01\%\\
 \hline
 \cite{Colaprete2010} & 3.0 g $\rm cm^{-3}$ & 155$^\dagger$ & \textemdash & 2500\textendash3800$^\dagger$ & 5.6 $\pm$ 2.9\% \\
 \hline
  \cite{Strycker2013} & 3.0 g $\rm cm^{-3}$ & 141$^{\dagger\dagger}$ & 2110 & 1840\textendash2640 & 6.3 $\pm$ 1.6\%\\
 \hline
 \end{tabular}\\
 \caption{LCROSS debris plume masses derived from stratified lunar sediment models at a water ice albedo of 0.8 versus previous studies \\
 $^\dagger$ These masses for water and the total plume were observed by the Shepherding Spacecraft and were temporal maxima that occurred at different times. The water includes vapor and ice.\\
 $^{\dagger\dagger}$ These data from \cite{Colaprete2010} include vapor and ice averaged over times 0--23 s post-impact.}
 \label{t:plume}
 \end{center}
\end{table}

Table \ref{t:plume} gives the masses of water ice and regolith and the concentration of water ice in the debris plume for the Three Layer, Decreasing Ice, and Increasing Ice models for the previously considered range of the albedo of pure water ice, $A_{W}$, since $x$ depends on $A_{W}$. The total debris plume masses presented here agree with previous results. For the Three Layer and Decreasing Ice models, the values for mass and mass concentration of water ice present within the debris plume are slightly larger than but still comparable to those derived from the spectroscopic measurements in \cite{Colaprete2010}; for the Increasing Ice model, the values for mass and mass concentration of water ice within the debris plume agree with those derived in \cite{Colaprete2010}. The total masses of the debris plume for all models agree with the range of masses presented by \cite{Colaprete2010}, and with predictions by \cite{Schultz2010} that the plume material ejected to a height of at least 2 km above the crater floor would contain approximately the mass of the impactor, ~2000 kg. \\

\section{Conclusions}

We modeled the ejecta plume of the LCROSS impact using fixed, experimentally generated velocity and ejection angle distributions, and we used the vertical albedo profile of the lunar sediment as our variable parameter. The model that best fits the observations contains a layer of mixed regolith and water ice, with the ice concentration increasing with depth, and a bedrock layer with a tensile strength high enough to prevent excavation of material by the LCROSS impact that would have otherwise occurred. This model could be degenerate with a regolith with a particle radius that increases with depth; however, we have not included in our simulation any additional effects that such an increase in particle radius would have on density, initial velocity, or ejection angle, and therefore these results should not be considered to be an accurate simulation of a lunar regolith with an increasing particle radius. We calculated the concentration of water ice within the lunar sediment using our best fit albedo profiles and found that the observations are consistent with a pre-impact lunar sediment with a water ice concentration of 8.2 $\pm$ 0.01 \%wt, assuming a water ice albedo of 0.8. We also derived values for the mass and volume of water per square meter of lunar surface and the depths of each of the layers of lunar sediment for our best fit model, the model with ice concentration increasing with depth, and found that the observed plume is consistent with a pre-impact total mass of water ice within the permanently shadowed regions of Cabeus crater of $5 \pm 3 \times 10^{11}$ kg, and an average pre-impact mass of water ice of $8 \times 10^{11}$ kg for all other models.\\

We now consider whether the vertical albedo profiles from the best fit models provide insight into the delivery method and the history of the water ice within Cabeus crater. We discuss here two possible delivery mechanisms for the water ice: stochastic delivery by individual events, or implantation on the lunar surface by the solar wind and eventual capture by the cold trap of the PSR. In the stochastic delivery scenario, water would be delivered during one or more individual events and should be found in layers throughout the crater floor material, though these layers may not have distinct boundaries due to mixing. In the solar wind implantation scenario, solar wind protons chemically bind to oxygen within lunar surface rocks to form OH and $\rm H_{2}O$ \citep{Crider2000, Crider2002}. They then ``hop'' across the lunar surface until they happen to land in a PSR, where they are cut off from the sunlight and no longer participate in the lunar water cycle. The water ice particles within the PSR may then vertically migrate deeper into the lunar regolith. A diurnally varying OH/$\rm H_{2}$O signature on the lunar surface, which supports the existence of such a cycle, was detected by the Moon Mineralogy Mapper ($\rm M^{3}$) instrument on the Chandrayaan-1 mission and by the Deep Impact Spacecraft \citep{Pieters2009, McCord2011, Sunshine2009}. For both methods, the water ice near the surface may be disturbed by meteorite gardening.\\

The best fit albedo profile includes a layer with water ice concentration increasing with depth, which does not rule out either of the two delivery methods proposed above. Because we are not sensitive to the first few centimeters of material, we cannot determine whether there is a thin layer of frost at the surface or if shallow layers of regolith dredged up from micrometeorite impacts covered older layers of water ice. We are limited here both by the vertical resolution of our models, which are most sensitive to layers of material at least two meters below the surface, and by the vaporization of the first couple of centimeters of water ice during the impact \citep{Stopar2018}.\\

One possible individual event that could have been a source of the water in Cabeus crater is water rich material deposited during a period of volcanic outgassing about 3.5 Gyr ago \citep{Needham2017}. The increasing ice model may be consistent with a deep, ancient water-rich layer or with a solar wind deposited layer that has been mixed and disrupted such that the ice concentration is higher deeper below the surface. Such an ancient layer would be consistent with delivery by volcanism, by impacts during the Late Heavy Bombardment, or by solar wind implantation followed by impact gardening. We are only able to differentiate between separated, non-adjacent water ice layers. Recent studies have found that multiple sources have likely contributed to the water ice reservoirs within lunar polar cold traps at various points throughout lunar history \citep{Deutsch2020}. If an ancient layer or a layer deposited by the original impactor were to lie underneath a layer of regolith mixed with water captured by the PSR cold trap with no layer of pure regolith in between, we would not be able to distinguish the layers from one another. We therefore cannot rule out the possibility that there may be multiple delivery sources, and we cannot distinguish between the possible delivery scenarios using the vertical albedo profile that best fits the observed LCROSS debris plume lightcurve.\\

We found that the observed debris plume is consistent with a layer of mixed regolith and water ice at the LCROSS impact site that is $5.8 \pm 0.75$ meters thick, with no additional layer of dry regolith. We did not find that the data supported a choice between models that differ only by the behavior of the mixed water ice and regolith; however, a model with ice concentration that increases with depth results in total ice mass and mass concentrations that agree with those derived by spectroscopic observations \citep{Colaprete2010}. The ground-based observations are consistent with an impact that excavated down to bedrock or another material with a high tensile strength at a depth of $5.8 \pm 0.75$ meters. Our best fit model includes an increasing ice concentration with depth, followed by a transition into a ``bedrock'' layer with material of sufficient hardness to prevent excavation; we note that a mixture of cold regolith and ice would match this description. Moreover, the increasing ice concentration up to the ``bedrock'' cutoff depth may support the idea that the ice concentration reached a level sufficient to cause the material to resist excavation. This physical interpretation of the ``bedrock'' layer suggests that additional ice may be present in Cabeus crater that was not excavated by the LCROSS impact, and that was therefore not accounted for in our calculations of water ice mass, volume, and concentration. Assuming a regolith density of 1.5 g $\rm cm^{-3}$, we calculated a water ice concentration within pre-impact lunar sediment within permanently shadowed regions of Cabeus crater of 8.2 $\pm$ 0.01 \%wt and a maximum mass of water ice within the entire crater of $5 \pm 3 \times 10^{11}$ kg. The pre-impact sediment properties derived by fitting our stratified lunar sediment models to the ground-based observations of the LCROSS debris plume are consistent with values inferred using other methods, and offer a complementary perspective on the lunar sediment excavated by the LCROSS impact. \\ 

\section{Acknowledgements}
We thank Charles Miller for helpful contributions and discussions during the development of the model and during the lightcurve analysis process. We acknowledge crucial work done by Fred Smalley on the debris plume model. We also thank Anthony Colaprete for helpful conversations about the spectroscopic results from the LCROSS Shepherding Spacecraft, including insight into the average lifetime of a water ice particle within the plume. Finally, we thank two anonymous reviewers, whose comments substantially improved this manuscript. This work was supported by NASA's Lunar Data Analysis Program through grant number NNX15AP92G.

\end{document}